\documentstyle[twocolumn,pre,aps,epsf]{revtex}

\newcommand{\be}{\begin{equation}}
\newcommand{\ee}{\end{equation}}
\newcommand{\bear}{\begin{eqnarray}}
\newcommand{\eear}{\end{eqnarray}}
\def\Tc{{T_{\rm c}}}
\def\kDOS{{\bf k}{\rm DOS}}

\begin{document}

\twocolumn[\hsize\textwidth\columnwidth\hsize\csname @twocolumnfalse\endcsname
\title{Enhancement of superconductive critical temperatures in almost empty or
full bands in two dimensions: possible relevance to $\beta$-HfNCl, C$_{60}$ and
MgB$_2$}

\author{Mahito Kohmoto${}^1$, Iksoo Chang${}^2$ and
Jacques Friedel${}^3$
}                     

\address{${}^1$Institute for Solid State Physics,
University of Tokyo, 5-1-5 Kashiwanoha, Chiba 277-8581, Japan}
\address{
${}^2$ Department of Physics, Pusan National University, Pusan
609-735, Korea}
\address{ ${}^3$Laboratoir\`{e} de Physique des Solides, Universit\'{e}
Paris-Sud,
   Centre d'Orsay, 91405 Orsay Cedex, France, unit associated to the CNRS}

\maketitle
\begin{abstract}
We examine possibility of enhancement of superconductive critical
temperature in
   two-dimensions.
The weak coupling BCS theory is applied, especially when
the Fermi level is near the edges of the electronic bands.
The
attractive interaction depends on ${\bf k}$ due to screening. The density of
states(DOS) does not have
a peak near the bottom of the band, but
$k$-dependent contribution to DOS (electron density on the Fermi surface) has a
diverging peak at the bottom or top. These features  lead to
significant enhancement of the critical temperatures. The
results are qualitatively consistent with the superconductive behaviors
   of HfNCl ($\Tc \le 25K$) and ZrNCl($\Tc
\le 15K$), C$_{60}$ with a field-effect
transistor configuration  ($\Tc = 52K$), and MgB$_2$ ($\Tc
\approx 40K$)  which have the unexpectedly high
  critical  temperatures.

\end{abstract}

]

\narrowtext

\section{Introduction}
There are much interests in the anisotropic superconductors. It is definitely
influenced
    by the  discovery  of  the  high $T_c$ cuprates
superconductors in which CuO$_2$ planes are considered to play an
essential role.\cite{BM}. Many properties are not simply described by the
standard
classical BCS theory\cite{BCS}.

Recently a number of compounds  having unexpectedly high
superconductive critical temperatures ($\Tc$) without CuO$_2$ planes  were
found.
These include layered nitride
(ZrNCl($\Tc
\le 15K$ \cite{zr},HfNCl ($\Tc \le 25K$) \cite{hf} , C$_{60}$ with a
field-effect
transistor configuration ($\Tc = 52K$) \cite{c60}, and MgB$_2$ ($\Tc
\approx 40K$) \cite{mg}. The common feature of these superconductors  is
notable anisotropy. Therefore it is mandatory to scrutinize the effects of
two-dimensionality on superconductivity.

In this paper we address a question of the role of quasi two-dimensionality in
    superconductivity.  We choose a model of   two-dimensional electronic band.
   Significant enhancement of $\Tc$  was
obtained when the band is almost empty or full. This result does not
qualitatively depend on the details of the model.

\section{Density of states and electron density of the two dimensional band  }

The  band in two dimensions is often
modeled by

\begin{equation}
\varepsilon({\bf k}) \simeq -2 t (\cos k_x + \cos k_y ),
\end{equation}
where $t$ is the transfer integral of the tight-binding model on the
square lattice.
This band has the  van Hove singularity in the density of
states(DOS) at half-filling.  There are a number of
works\cite{hs,friedel,newns,abrikosov,bok,fk,cfk} that try to explain high
$T_c$ cuprates by the van Hove singularity.

On the other hand, here we focus on the van Hove
singularity at the edge of band, namely  a nearly empty band (or an almost full
band) is considered. The density of states(DOS) does not diverge, but the
${\bf k}$-dependent density of states $\kDOS$ (electron density at the Fermi
surface)  diverges at
$\Gamma$ (${\bf k}=0$) (the bottom of the band) and ($\pm\pi, \pm\pi$) (the top
of the band). See Fig. 1. This is the key difference from the case for a
degenerate semiconductor\cite{cohen}  which is three dimensional and  does not
have a van Hove singularity at the band edges. The density of states is
given by
\begin{equation}
N(E)=\int \frac{1}{|\bigtriangledown
\varepsilon({\bf k})|} dl = \int \frac{1}{v} dl,
\label{DOS}
\end{equation}
where $l$ is the Fermi surface (line in two dimensions) and $v$ is the
semiclassical
velocity. Here

\be
\kDOS =1/|\bigtriangledown
\varepsilon({\bf k})|
\ee
and it is plotted in Fig. 1.
Note that $\kDOS$ is diverging near the $\Gamma$ point
${\bf k}=(0,0)$ as well as at ($\pm \pi, \pm \pi$) and
at $(\pm\pi,0), ~(0,\pm\pi)$.

\begin{figure}

\narrowtext
\hbox{\epsfysize=2.3in \epsffile{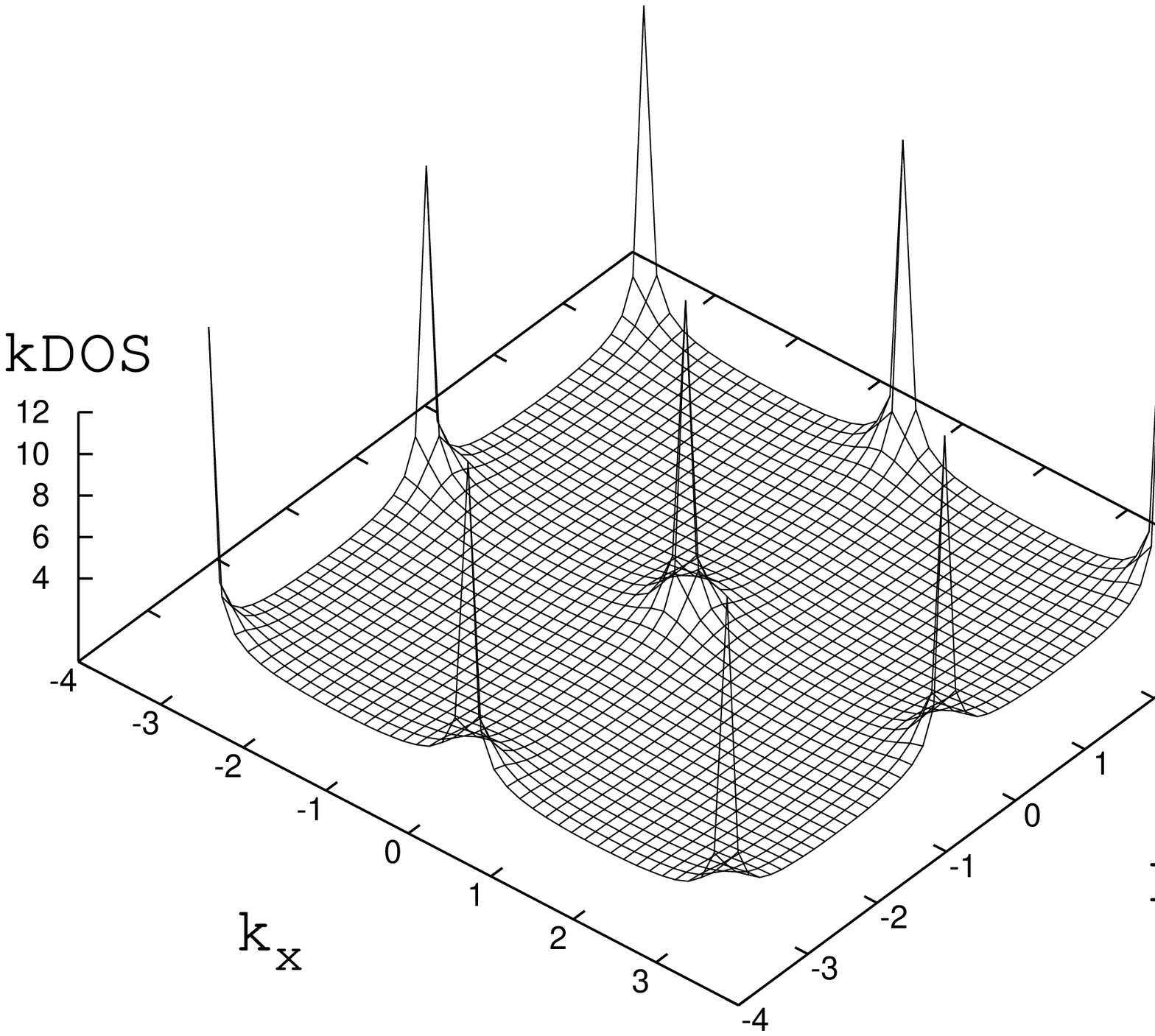}}
\caption{
$\kDOS$. All the peaks are actually
diverging.
The peak at $\Gamma (0,0)$ corresponds to  the
bottom of the band . The four peaks at ($\pm \pi, \pm \pi$) are at the
top of the
band and the others give van Hove singularities at half-filling.
}
\end{figure}

These singularities are not seen in DOS since it is integrated on a vanishingly
short
Fermi line. Instead DOS is almost constant near the band edges. See Fig. 2.

\begin{figure}

\narrowtext
\hbox{\epsfysize=2.3in \epsffile{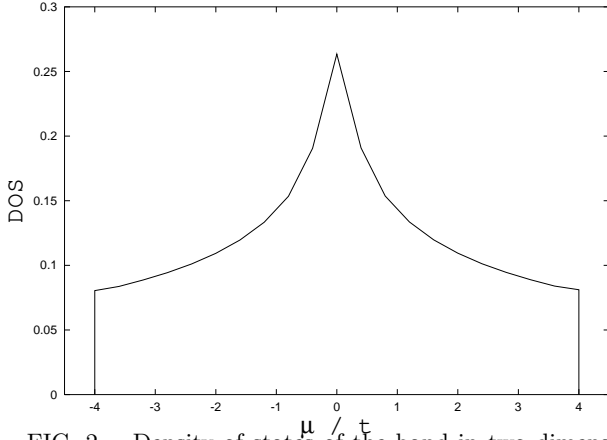}}
\caption{
Density of states of the band in two dimensions given by Eq.(1). The peak at
the center is actually diverging }
\end{figure}

These
behaviors near the band edges are rather universal  since the band
dispersion may
be approximated by  $E_{edge} \pm  (k_x^2 + k_y^2)/2m^*$ in general
including that of the honeycomb lattice which is relevant to HfNCl( and ZrNCl)
and MgB$_2$ where $m^*$ is the effective band mass, $E_{edge}$ is the band
edge, and
$+$ and
$-$ apply to near the bottom and top of the band respectively. The
important point
is that if the interactions depend on
${\bf k}$,
$\kDOS$ has to be considered  carefully.

\section{BCS formula and screening }

With the properties of the band above in mind, we consider the gap equation,

\begin{equation}
\Delta_k = -
\sum_{k'}\frac{V_{kk'}\Delta_{k'}(T)}{2E_{k'}}\tanh\frac{E_{k'}}{2k_B T},
\label{gap}
\end{equation}
where $E_k = \sqrt{\{\varepsilon(k) -\mu\}^2 +\Delta_k (T) ^2}$,
$\Delta_k(T)$ is
the gap
order parameter and $V_{kk'}$ is the interaction. The sum is restricted
within the
cutoff
$\mu -E_c<\varepsilon(k')<\mu+E_c$ where $\mu$ is the chemical potential.

   From the gap equation(\ref{gap}) one can obtain $T_c$ and $\Delta (T)$.
Near $T_c$,
$\Delta$ is very small; then (\ref{gap}) is linearized. $T_c$ is determined
by the
linearized equation. The gap $\Delta(T)$ is obtained by iteration of
(\ref{gap}).

In the usual BCS interaction one takes $-V$ a constant if  two
particles are
both within the cutoff and vanishes otherwise. This implicitly assumes
point-like
electron-electron interactions, namely the screening is perfect. In this
case the
classical BCS result is

\begin{equation}
T_c
\sim 1.13E_c \exp (-1/NV),
\label{bcs}
\end{equation}
   where N is the DOS near the bottom of the band.
When one
takes physically reasonable value of $V$, enhancement of $T_c$ can not be
expected
even if $\kDOS$
is large. This is because the Fermi line is a small circle and the range of
integration is very small. It offsets the large but uniform $\kDOS$. This
behavior is
totally analogous to the fact that the behavior of DOS --integral of $\kDOS$
along the
Fermi line-- that is not enhanced near the bottom of the band. See Fig. 2.

On the other hand if the
interaction depends on
${\bf q}={\bf k}-{\bf k'}$, {\it i.e.} is not a
constant in the integral, the effect of the
large $\kDOS$ is not necessarily  canceled by the short length of the Fermi
line. Specifically we choose an interaction
which is screened and has a peak at ${\bf q}={\bf k}-{\bf k'}=0$ :
\cite{deGennes},

\begin{equation}
V_q = - \frac{g_q^2}{q^2 +q_0 ^2}.
\label{phonon}
\end{equation}
Here $g_q$ is the coupling constant and $q_0$ is the inverse of the screening
length $L$: $q_0 \simeq  1/L$.
The screening length actually depends
on $\mu$, but we neglect this
effect for the sake of simplicity.
The attractive interaction
   (\ref{phonon}) depends on
$k$ and has a peak at
${\bf q}=0$. Thus the effect of the large $\kDOS$ can not be totally
canceled by
the small Fermi line, as discussed above.

\section{Effective interaction }
Let us first give an estimate of the effective interaction. For the chemical
potential $\mu = -4t+
\eta$
where $-4t$ is the energy of the bottom of the band in the noninteracting case
and
$\eta$ is small but larger than the cutoff
$E_c$, one has, from (\ref{gap}),

\begin{eqnarray}
2 &= & \nonumber \\
&&\int_{0}^{2\pi} \int_{k'_{inf}}^{k'_{sup}}
\frac{g_q ^2  \tanh (\frac{k'^2t-\eta}{2k_BT_c})
}{k'^2 + k^2-2k'k\cos\theta + q_0^2}
~\frac{k'dk'd\theta}{k'^2t-\eta}
\end{eqnarray}
with $k^2t = \eta, \eta \gg E_c$ and $|k'^2t - \eta|<E_c$.
This leads to
\be
2=\pi g_q^2t \int  \tanh (\frac{ \kappa}{2k_BT_c})~\frac{d\kappa}{\kappa} {1
\over A}
\ee
with $k'^2t-\eta = \kappa$ and $A=\kappa^2+4\eta ~q_0^2 t+\cdots$. With the
preceding conditions, and if $\eta$ is large enough versus $E_c$, $A$
reduces to $1
/ 4\eta ~q_0^2 t $. One reverts to the BCS case of $V$ constant but with

\be
V_{kk'} = -\frac{g_q^2}{q_0^2},
\ee
thus  higher $T_c$ is expected. This result is also apply when the Fermi energy
is near the top of the band (hole doped) as seen by changing the signs of
$t$ and
$\eta$.

\section{Numerical results }

The numerical solution is consistent with the above analysis and gives
enhancement of critical temperatures near the bottom of the band. They are
plotted in Fig. 3 for a number of screening lengths and cutoffs. They
take the  maximum values at $\mu_{op}$ near the bottom of the band. The
decrease of
$T_c$'s for
$\mu$ smaller than
$\mu_{op}$  is due to the semiconductor gap where DOS vanishes. But note
that $T_c$'s remain finite at the bottom of the band in the noninteracting
case.
They extend to the band gap
region due to the superconductive coherence effect. It is natural that
$T_c$'s do not
depend on the cutoff in this region. The transition temperature
$T_c$ also decreases for
$\mu$ larger than  $\mu_{op}$. This is because the $\kDOS$ is getting smaller
in this
region. Note that if screening is poorer, $Tc$ is higher.

\begin{figure}

\narrowtext
\hbox{\epsfysize=2.3in \epsffile{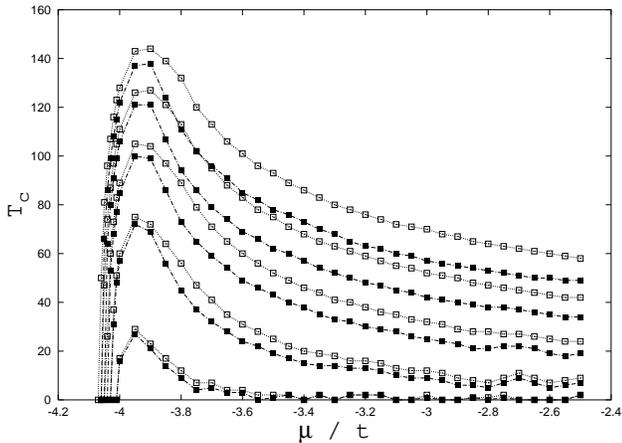}}
\caption{
Critical temperature as a function of $\mu/t$.
The total band width is $8t$ and $-4t$ is the bottom of band. The
parameters are:
transfer
$t=0.25$eV, screening lengths $L=5, 10, 15, 20, 25$
lattice spacings respectively for the lower to higher curves
and $g_q^2=0.6t$.
The cutoffs $E_c$ are $30$meV (an occupied square mark) and
$50$meV (an empty square mark).
}
\end{figure}

\begin{figure}

\narrowtext
\hbox{\epsfysize=2.3in \epsffile{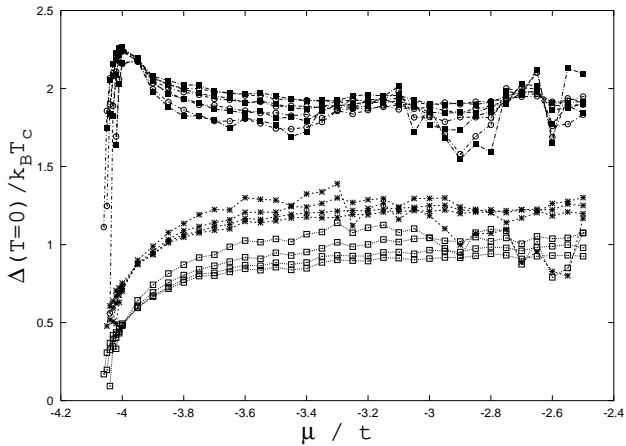}}
\caption{
The upper set of curves are  the ratios $\Delta(0)/T_c$ for
the maximum of $\Delta(0)$'s.
Within this set, the curves from lower to higher ones correspond to
screening lengths $L=10, 15, 20, 25$ lattice spacings
for the cutoff energy
$E_c = 30$ meV by an occupied square mark,
$E_c = 50$ meV by an empty circle mark.
The lower set of curves are  the ratios $\Delta(0)/T_c$ for the
minimum of $\Delta(0)$'s . Screening lengths are $L=10, 15, 20, 25$ lattice
spacings for the higher to lower curves respectively for the cutoff energy
$E_c = 30$ meV by a star mark,
$E_c = 50$ meV by an empty square mark.
}
\end{figure}

We have a typical
$s$-wave pairings, but the gaps depend on
the absolute value of $k,|{\bf k}|$. The maximum and minimum of
$\Delta(0) / T_c$'s are plotted in Fig. 4. Note that the maxima of the gaps are
almost independent of the cutoff.
They are about $1.8 Tc$ which is close to the BCS value
$1.76 \Tc$ at $\mu \simeq -3.4$. They  increase as $\mu$ is lowered. On the
other
hand the
minima of the gap depend on the cutoff as expected, since the minima
occur at
the edge of the cutoff. They are less than the BCS value and decrease as
$\mu$ is
decreased. However, less anisotropic $s$ gap is possible in cases with shorter
screening lengths\cite{cfk}.

   Note that
$\Delta(0)/ T_c$'s approach zero in a similar manner to the $T_c$'s.
This implies
that
$\Delta(0)$ decay faster than $\Tc$.

\section{Medium to high $\Tc$ anisotropic superconductors}
In this Section we briefly discuss the known superconductive properties of
some of the  anisotropic compounds which have unexpectedly high critical
temperatures. Most of the properties are consistent with our results
qualitatively.

\subsection{  ZrNCl and HfNCl }

It is known that a series of transition-metal nitrides with the rock-salt
structure become superconductors, such as TiN, ZrN, HfN and NbN with $\Tc$s at
5.5, 10.7, 8.8 and 18.0K, respectively \cite{nit}.  Among these
$\beta$-ZrNCl is
characterized as a semiconductor with a band gap of $\sim 3$eV. On lithium
intercalation  the electrons are
transferred from the intercalated lithium atoms to the ZrN layers through
chlorine layers, and the empty $t_{2g}$ band is partially filled with
electrons,
giving the metallic behavior. It becomes a superconductor with a $\Tc$ of $\le
15$K. It is suggested that the superconductivity occurs within the thin
two-dimensional ZrN layers separated by the close-packed chlorine layers
\cite{zr}.

$\beta$-HfNCl is isostructual with $\beta$-ZrNCl. However, expanded spacings
indicate the formation of co-intercalated phases of THF molecules with lithium
between the layers \cite{hf}. The much higher $\Tc \le 25$K of
Li$_{0.48}$(THF)$_y$HfNCl\cite{hf} than that of the zirconium analogue 
($\Tc \le
15K$)\cite{zr} may be attributed to more prominent two-dimensionality due
to the
co-intercalation. Note that $\Tc =8.8K$ of HfN the rock-salt structure is
lower
than that of ZnN (=10.7K). Recently it has been shown that a larger interlayer
spacing leads to weaker screening within a layer\cite{sk}. This result may
explain the above phenomenon.

\subsection{ C$_{60}$ }

A crystal of C$_{60}$ molecules superconducts when doped with alkali metal
atoms,
which donate electrons to the C$_{60}$ lattice \cite{c601,c602}. The critical
temperature range from below 10K to 33K, depending on dopant. Recently,
Sch$\ddot{o}$n et al.
\cite{c60} succeeded in making a field-effect transistor( FET) device on a
C$_{60}$ crystal. This enables to adding holes or electrons into the top
layer of
a C$_{60}$ crystal. Doping level is continuously varied by means of the applied
gate voltage, without the crystallographic changes. The peaked behaviors of
$\Tc$
are observed in both electron and hole doped cases. The peak value in the
hole-dope case is remarkably high $52$K. That in the electron-dope case is
$11$K.
The result was discussed in relation to the band structure calculation for the
bulk three dimensional solid
\cite{band}. The Fermi energy is not near the 3D band edge. However, the
energies of the surface states which are pertinent to the present case have
tendency to be in the energy gaps of the bulk states. The present
situation of the two-dimensional superconductivity requires appropriate new
calculations.

\subsection{  MgB$_2$}

The critical temperature of MgB$_2$ $\Tc \sim 40$K is by far the highest in any
binary compound\cite{mg}. Boron forms layers of honeycomb lattices with
magnesium
as a space filler. A number of band structure calculations are
available\cite{band1,band2,band3,band4}. There is no major disagreement in
these
calculations and indicate that Mg is substantially ionized. The bands at the
Fermi energy derive mainly from B orbitals. These are the bonding ($p_{x,y}$)
$\sigma$ bands (2D like), the bonding ($p_z$) $\pi$ bands(3D like) and the
antibonding ($p_z$)
$\pi$ bands(3D like).

There is a close analogy between MgB$_2$ and graphite. They have the same layer
structure. Graphite is isoelectronic with MgB$_2$. The 2D-like $\sigma$ band of
graphite is completely filled and it is not related to the superconductive
instability. Graphite becomes superconducting up to $5$K, but only when
electron-doped (intercalated)\cite{graphite}. On the other hand MgB$_2$
superconducts at stoichiometry at $\Tc \sim 40$K. This comparison strongly
suggests that 2D-like
$\sigma$ band plays an important role in  MgB$_2$. The Fermi energy is near
the top of this band. This situation corresponds to the case we
investigated above
for enhancement of $\Tc$. The other states on the 3D-like bands become
superconductive due to interband sctterings\cite{suhl,kondo}.

\section{Conclusions}

We study the superconductive properties of the two-dimensional bands which is
almost full(hole-doped) or empty(electron-doped).
   The crucial and specific point is the large
electron density $\kDOS$ near the top or the bottom of the bands. It is shown
that  screened interaction leads to  enhancement of   superconductive critical
temperature
$\Tc$ from the approximate  analysis and the numerical computations. The poorer
the screening, the stronger the  enhancement of $\Tc$ is.

Since the physical origin of the enhancement of $\Tc$ is rather clear (large
electron density), the strong coupling treatment which might be more
appropriate
for some  compounds will not likely to change the results qualitatively.

The present  results are not in conflict  with most of the known
superconductive properties of some of the
  anisotropic compounds which have unexpectedly high critical
temperatures such as $\beta$-HfNCl, C$_{60}$ and
MgB$_2$. \\

\begin {center}
{\large{\bf Acknowledgments}}
\end {center}

It is a pleasure to thank M. Takigawa for a useful discussion.
We are grateful to Korea-Japan binational
program through KOSEF-JSPS.




\begin{thebibliography}{}

\bibitem{BM}  L.G. Bednorz and K.A. M\"{u}ller, Z. Phys. B{\bf 64}, 199 (1986).

\bibitem{BCS} J. Bardeen, L.N. Cooper, J.R. Schrieffer, Phys. Rev. {\bf
106}, 162;
{\bf 108}, 1175 (1957).

\bibitem{zr} S. Yamanaka, H. Kawaji, K. Hotehama, M. Ohashi, Adv. Matter. {\bf
8}, 771 (1996).

\bibitem{hf} S. Yamanaka, K. Hotehama,H. Kawaji, Nature {\bf
392}, 580 (1998).

\bibitem{c60} J.H. Sch$\ddot{o}$n, Ch. Kloc and B. Batlogg, Nature {\bf
408}, 549 (2000).

\bibitem{mg} J. Akimitsu, Symposium on Transition Metal Oxides, Sendai, January
10, 2001; N. Nakagawa, T. Muranaka, Y. Zenitani, and J. Akimitsu, Nature {\bf
410}, 63 (2001).


\bibitem{muon} G.M. Luke, Y. Fujimoto, K.M. Kojima, M.I. Larkin, J. Merrin, B.
Nachumi, Y.J. Uemura. Y. Maeno, Z.Q. Mao, Y. Mori, H. Nakamura and M. Sigrist,
Nature {\bf 394}, 558 (1998).


\bibitem{hs} J.E. Hirsch and
D.J. Scalapino, Phys. Rev. Lett. {\bf 56}, 2732 (1986).

\bibitem{friedel} J. Friedel, Physica C, {\bf 153-5}, 1610(1988); J.
Phys: Condens. Matter. {\bf 1}, 7757 (1989); Nato Institute on Condensed
Matter,
Biarritz (1990).

\bibitem{newns} D.M. Newns, C.C. Tuei, P.C. Pattnaik and C.L. Kane, Comments
Cond.
Mat. Physics {\bf 15} 273 (1992).

\bibitem{abrikosov} A.A. Abrikosov, Physica C{\bf 222}, 191 (1994); {\it 
inbid}.
C{\bf 244}, 243 (1996).

\bibitem{bok} J. Bouvier and J. Bok, Physica C {\bf 249}, 117 (1995); '{\it Gap
Symmetry and Fluctuations in High $T_c$ Superconductors}' ed. J. Bok, G.
Deutscher
, D. Pavunaand and S.A. Wolf, Plenum Press, New York (1998)  NATO ASI Series,
Series B Physics {\bf 371} 77 .


\bibitem{fk} J. Friedel and M. Kohmoto, to be published in Int. J. Mod.
Phys. B (2001).

\bibitem{cfk} I. Chang, J. Friedel and M. Kohmoto, Europhys. Lett. {\bf 50},
782 (2000).

\bibitem{cohen} M.L. Cohen  {\it in ``Superconductivity"} ed. Parks,
Vol.1, Chapter 12.  (Dekker, New York, 1969). p.615.


\bibitem{deGennes} see, for example, P.G. de Gennes, {\it Superconductivity in
metals an alloys} (Benjamin, New York, 1966) Chapter 4.

\bibitem{nit} B.W. Roberts,  J. Phys. Chem. Ref. Data, {\bf 5}, 581
(1976);

N.Pesall and J.K. Hulm, Physics {\bf 2}, 311 (1966).

\bibitem{zr} H. Kawaji, K. Hotehama, S. Yamanaka, Chem. Matter {\bf 9}, 2127
(1997).

\bibitem{sk} H. Shimahara and M. Kohmoto, cond-mat/0103402.

\bibitem{c601} M. J. Rosseinsky et al., Phys. Rev. Lett. {\bf 66}, 2830 (1991).

\bibitem{c602} K. Holczer, O. Klein, S.-M. Huang, R.B. Kaner, K.-J. Fu, R.L.
Whetten, F. Diederich, Science {\bf 252}, 1154 (1991).

\bibitem{band} S.C. Erwin, in {\it Buckminsterfullerens} ed. W.E. Billups and
M.A. Ciufolini, 217 (VCH, New York, 1992).

\bibitem{band1} J. Kortus, I.I. Mazin, K.D. Belashchenko, V.P. Antropov, L.L.
Boyer, cond-mat/0101446.

\bibitem{band2} K.D. Belashchenko, M. van Schilfgaarde and V.P. Antropov,
cond-mat/0102290.

\bibitem{band3} G. Satta, G. Profeta, F. Bernardini, A. Continenza, S.
Massidda,
cond-mat/0102358.

\bibitem{band4} J.M. An and W.E. Pickett, cond-mat/0102391.

\bibitem{graphite} I.T. Belash et al., Solid State Commun. {\bf 64}, 1445
(1987).

\bibitem{suhl} H. Suhl, B.T. Mattihias, and L.R. Walker, Phys. Rev. Lett. {\bf
3}, 552 (1959).

\bibitem{kondo} J. Kondo, Prog. Theor. Phys. {\bf 29}, 1 (1963).





\end{thebibliography}
\end{document}